\documentclass[aps,prd,showpacs,preprintnumbers,twocolumn]{revtex4}
\usepackage[dvips]{graphicx}
\usepackage{bm,latexsym,amsmath,amssymb,amsfonts}
\begin{document}
\title{Relativistic stars in $f(R)$ gravity, and absence thereof}

\author{Tsutomu~Kobayashi$^{1}$}
\email[Email: ]{tsutomu"at"gravity.phys.waseda.ac.jp}
\author{Kei-ichi~Maeda$^{1\,,2}$}
\email[Email: ]{maeda"at"waseda.jp}
\address{\,\\ \,\\
$^{1}$ Department of Physics, Waseda University,
Okubo 3-4-1, Shinjuku, Tokyo 169-8555, Japan\\
$^{2}$ Advanced Research Institute for Science and Engineering,
 Waseda University,
Okubo 3-4-1, Shinjuku, Tokyo 169-8555, Japan}

\begin{abstract}
Several $f(R)$ modified gravity models have been proposed which realize
the correct cosmological evolution
and satisfy solar system and laboratory tests.
Although nonrelativistic stellar configurations can be constructed,
we argue that relativistic stars cannot be present in such $f(R)$ theories.
This problem appears due to the dynamics of the effective scalar degree of freedom
in the strong gravity regime.
Our claim thus raises doubts on the viability of $f(R)$ models.
\end{abstract}

\pacs{04.50.Kd, 04.40.Dg, 95.36.+x}
\preprint{WU-AP/289/08}
\maketitle

\section{Introduction}

The current accelerated expansion of the Universe
is one of the deepest mysteries in cosmology~\cite{Acceleration}.
This acceleration may be due to some unknown energy-momentum component
having the equation of state $p/\rho \approx -1$,
or may be due to a modification of general relativity.
In this paper, we are interested in the latter possibility.
The simplest phenomenological way of modifying gravity is
to consider a gravitational action described by a function of the Ricci scalar, $f(R)$,
instead of the Einstein-Hilbert action.
An early attempt is found, e.g., in~\cite{StInf},
where a modification like $f(R)= R+R^2/\mu^2$ was used
to explain the accelerated expansion in the {\em early} Universe.
More recently, $f(R)$ modified gravity theories are
often considered as a possible origin of the {\em current} acceleration of
the Universe~\cite{Rev}.

Any modified theories of gravity must account for
the late-time cosmology which is well established by observations,
and at the same time must be consistent with
solar system and laboratory tests of gravity.
However,
since $f(R)$ gravity has the equivalent description
in terms of the Brans-Dicke theory with the Brans-Dicke parameter
$\omega=0$~\cite{BD},
naively constructed models would result in
violation of the above requirements~\cite{Chiba, ESK, CSE}.
For example, the original proposal of~\cite{CDTT} employs
$f(R) = R-\mu^4/R$, which admits an acceleratedly expanding solution
even in the absence of a dark energy component.
In order to accommodate a late time acceleration, however,
one must introduce a very low energy scale, $\mu\sim H_0$ (the present Hubble 
scale),
leading to a very light scalar field,
which predicts the parametrized post-Newtonian (PPN)
parameter $\gamma=(1+\omega)/(2+\omega)=1/2$. Obviously,
this result contradicts the observational constraint $|\gamma-1|\lesssim 
10^{-4}$~\cite{Will}

To circumvent this difficulty, it is important to notice that
the presence of matter may affect the dynamics of the extra scalar degree 
of freedom.
The key idea is essentially the same as that of the ``chameleon'' 
model~\cite{Chameleon, Cham2, Cham3},
in which the effective mass of the scalar field depends on the local matter
 density.
In particular, the scalar field is very light for the cosmological density
and is heavy for the solar system density, 
though the actual mechanism is 
slightly more complicated.
The most successful class of $f(R)$ models~\cite{Cembranos:2005fi, Hall, NvA, Tegmark, Li, AT, St, Hu, AB} 
incorporates this
chameleon mechanism to evade local gravity tests.
The experimental and observational
consequences of this kind of $f(R)$ models are found in 
Refs.~\cite{Tsujikawa1, Tsujikawa2, Brax}.
(See also~\cite{NO}.)

In this paper, we consider the strong gravity regime of
the carefully constructed models of~\cite{St, Hu, AB}.
The strong gravity aspects of $f(R)$ theories
have not been explored so much before.
Recently, Frolov suggested that
such $f(R)$ models generically suffer from the problem of curvature 
singularities
which can be easily accessed by the field dynamics {\em in the presence of 
matter}~\cite{Frolov}.
In other words, a curvature singularity may be caused not by
diverging gravitational potential depth, $|\Phi|=\infty$, but rather by
a slightly large gravitational field, $|\Phi|\lesssim 1/2$.
This motivates us to study relativistic stars in the context of $f(R)$ gravity.
Spherically symmetric stars in $f(R)$ gravity have been investigated so far
 in~\cite{Stars1, Stars2, MV, BB}. (We confine ourselves to a {\em metric}
 theory of $f(R)$ gravity. Using the {\em Palatini} formalism,
polytropic stars have been studied  in~\cite{Sot}.)
We shall show, both analytically and numerically, that
stellar solutions with relatively strong gravitational fields cannot be 
constructed.
Using the specific example of relativistic stars,
we clarify how the singularity problem arises in the strong gravity regime of $f(R)$ theories.
The singularity problem was also identified in~\cite{ABCos} in a cosmological setting.

This paper is organized as follows.
In the next section we derive equations of motion for $f(R)$ modified gravity,
and define the specific model we study. In Sec.~\ref{sec:basic},
we reinterpret the problem of finding the desired stellar configuration
as the problem of the particle motion in classical mechanics.
We give some analytic arguments in Sec.~\ref{sec:An} 
and then we present our numerical results in Sec.~\ref{sec:Num}.
Finally, we draw our conclusions in Sec.~\ref{sec:Conc}.

\section{Preliminaries}\label{sec:fr}

\subsection{$f(R)$ gravity}

The action we consider is given by
\begin{eqnarray}
S=\int d^4x\sqrt{-g}\left[\frac{f(R)}{16\pi G}+{\cal L}_{\rm m}\right],
\label{action}
\end{eqnarray}
where
$f(R)$ is a function of the Ricci scalar $R$, and
${\cal L}_{{\rm m}}$ is the Lagrangian of matter fields.
Variation with respect to metric leads to the field equations
\begin{eqnarray}
f_R R_{\mu\nu}-\nabla_{\mu}\nabla_{\nu}f_R
+\left(\Box f_R-\frac{1}{2}f\right)g_{\mu\nu}=8\pi G T_{\mu\nu},
\label{field_eq_original}
\end{eqnarray}
where $f_R:=df/dR$ and $T_{\mu\nu}:=-2\delta{\cal L}_{\rm m}/\delta 
g^{\mu\nu}+g_{\mu\nu}{\cal L}_{\rm m}$.
The trace of  Eq.~(\ref{field_eq_original}) reduces to
\begin{eqnarray}
\Box f_R=\frac{8\pi G}{3} T+\frac{1}{3}(2f-f_RR).\label{feqtr}
\end{eqnarray}
We now introduce an effective scalar degree of freedom by
defining $\chi:=f_R$.
Inverting this relation, the Ricci scalar can be expressed in terms of
$\chi$: $R=Q(\chi)$.
Thus, Eqs.~(\ref{field_eq_original}) and~(\ref{feqtr}) are equivalently 
rewritten as
\begin{eqnarray}
\chi G_{\mu}^{\;\nu}
&=&8\pi G T_{\mu}^{\;\nu}+\left(\nabla_{\mu}\nabla^{\nu}
-\delta_{\mu}^{\;\nu}\Box\right)\chi
-\chi^2 V(\chi)\delta_{\mu}^{\;\nu},
\label{fieldeq}
\\
\Box\chi&=&\frac{8\pi G}{3}T+\frac{2\chi^3}{3}\frac{dV}{d\chi},
\label{scalareq}
\end{eqnarray}
where the effective potential $V$ is given by
\begin{eqnarray}
V(\chi):=\frac{1}{2\chi^2}\left[\chi Q(\chi)-f(Q(\chi))\right],
\end{eqnarray}
and
\begin{eqnarray}
\frac{dV}{d\chi}=\frac{1}{2\chi^3}\left[2f(Q(\chi))-\chi Q(\chi)\right].
\end{eqnarray}

Eqs.~(\ref{fieldeq}) and~(\ref{scalareq}) are equivalent to
the Jordan frame equations of motion in the Brans-Dicke theory with $\omega=0$,
if we ignore the potential term $V(\chi)$.
One can move to the Einstein frame by performing the conformal transformation
$\tilde g_{\mu\nu} = \chi g_{\mu\nu}$ with $\chi = \exp (\sqrt{16\pi G/3}\,\phi )$,
where $\phi$ is the canonical scalar field.
The potential for $\phi$ is then given by $V(\chi(\phi))$.
Although the Einstein frame equations of motion are sometimes convenient,
we shall not use this and work in the Jordan frame directly throughout 
the paper, so as to
avoid confusion concerning the coupling between matter fields
and the effective scalar degree of freedom.

\subsection{The model}

\begin{figure}[tb]
  \begin{center}
    \includegraphics[keepaspectratio=true,height=56mm]{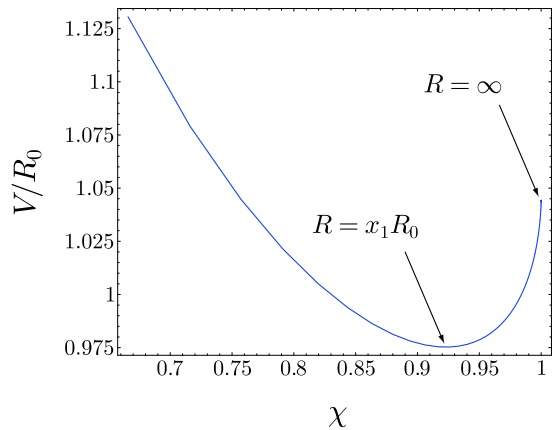}
  \end{center}
  \caption{The effective potential $V(\chi)$ for
  Starobinsky's $f(R)$ model with
  $n=1$ and $x_1=3.6$ ($\lambda\simeq 2$).
$\chi$ is the effective scalar degree of freedom defined by
$\chi :=df/dR$.}%
  \label{fig:v.eps}
\end{figure}

In order to be explicit,
we take $f(R)$ in the following form~\cite{St}:
\begin{eqnarray}
f(R)=R+\lambda R_0\left[\left(1+\frac{R^2}{R_0^2}\right)^{-n}-1\right],
\label{fRSt}
\end{eqnarray}
where $n, \lambda \,(>0)$, and $R_0 \, (>0)$ are parameters.
This model is carefully constructed so that it gives viable cosmology
and satisfies solar system and laboratory tests.
Suppose that the de Sitter solution in this theory is expressed as
$R=R_1=$ constant $=x_1 R_0$.
We may define the ``cosmological constant'' as $\Lambda_{{\rm eff}}:=R_1/4$.
Since the de Sitter solution follows from $dV/d\chi=0$, one has
\begin{eqnarray}
\lambda = \frac{1}{2}\frac{x_1(1+x_1^2)^{n+1}}{(1+x_1^2)^{n+1}-1-(n+1)x_1^2}.
\label{lambda=x}
\end{eqnarray}
Thus, we may use $x_1$ as a model parameter instead of $\lambda$. 
We will take as $x_1$ the maximal root of Eq.~(\ref{lambda=x}) for a given
$\lambda$, because
it corresponds to the de Sitter minimum of the potential.

The scalar field $\chi$ is written in terms of $R$ as
\begin{eqnarray}
\chi=1-2n\lambda\frac{R}{R_0}\left(1+\frac{R^2}{R_0^2}\right)^{-n-1}.
\end{eqnarray}
One sees that
$\chi\to1$ as $R\to\pm\infty$ and $R\to0$.
Note that curvature singularities, $R=\pm\infty$, is mapped to
the finite value of $\chi=1$.
The value of $\chi$ at the de Sitter minimum is given by
\begin{eqnarray}
\chi_1=1-\frac{nx_1^2}{(1+x_1^2)^{n+1}-1-(n+1)x_1^2}.
\end{eqnarray}
A typical form of the potential $V(\chi)$ around the de Sitter minimum is
shown in Fig.~\ref{fig:v.eps}.
In fact, the potential is a multivalued function of $\chi$,
and its shape is complicated, probably even pathological, away from the
 plotted region.
However, since our discussion here focuses on the behavior of $\chi$
around the de Sitter minimum, there is no difficulty with such a complicated 
potential.

Though we shall confine ourselves to the specific model defined by
 Eq.~(\ref{fRSt}),
the result will apply to other $f(R)$ models as well.
In particular, the models proposed by Hu and Sawicki~\cite{Hu}
and by Appleby and Battye~\cite{AB} fall into the same class as~\cite{St}
in the sense that the potential around the de Sitter minimum
has the same structure.

\section{Spherically symmetric stars in $f(R)$ gravity}\label{sec:basic}

\subsection{Basic equations}

To study stellar configurations in $f(R)$ gravity,
we take  the ansatz of a spherically symmetric and  static metric:
\begin{eqnarray}
ds^2=-N(r) dt^2+ \frac{1}{B(r)} dr^2+r^2\left(d\theta^2+\sin^2\theta
 d\varphi^2\right).
\label{met}
\end{eqnarray}
The energy-momentum tensor of matter fields is given by
\begin{eqnarray}
T_{\mu}^{\;\nu}=\text{diag}\left(-\rho, p, p, p\right).
\end{eqnarray}
From
the energy-momentum conservation, $\nabla_{\nu}T_{\mu}^{\;\nu}=0$, we obtain
\begin{eqnarray}
p'+\frac{N'}{2N}(\rho+p)=0.\label{dp=}
\end{eqnarray}
Here and hereafter a prime denotes differentiation with respect to $r$.
The $(tt)$ and $(rr)$ components of the field equations~(\ref{fieldeq}) yield,
 respectively,
\begin{eqnarray}
&&\frac{\chi}{r^2}\left(-1+B+rB'\right)=-8\pi G\rho-\chi^2V
\nonumber\\&&\qquad\qquad\qquad\qquad
-B\left[\chi''+\left(\frac{2}{r}+\frac{B'}{2B}\right)\chi'\right],
\label{fe_B}
\\
&&\frac{\chi}{r^2}\left(-1+B +rB\frac{N'}{N}\right)
=8\pi Gp-\chi^2V
\nonumber\\&&\qquad\qquad\qquad\qquad\qquad\quad
-B\left(\frac{2}{r}+\frac{N'}{2N}\right)\chi'.
\label{fe_N}
\end{eqnarray}
The equation of motion for $\chi$ [Eq.~(\ref{scalareq})] gives
\begin{eqnarray}
&&B\left[\chi''+\left(\frac{2}{r}+\frac{N'}{2N}+\frac{B'}{2B}\right)\chi'\right]
\nonumber\\&&\qquad\qquad
=\frac{8\pi G}{3}(-\rho+3p)+\frac{2\chi^3}{3}\frac{dV}{d\chi}.
\label{eq_chi}
\end{eqnarray}

We will not integrate the angular components of the field equations.
Instead, we will use them
to check the accuracy of our numerical results, because
those are derived from other equations via the Bianchi identity.

To specify the boundary conditions at the center of a star,
assuming the regularity,
we expand the variables in  the power series of $r$
as
\begin{eqnarray}
&&N(r)=  1+ N_2 r^2+... ,\;
B(r)= 1+ B_2 r^2+...,
\nonumber\\
&& \chi(r)= \chi_c\left(1+ \frac{C_2}{2}r^2+...\right),\label{series-r=0}
\\
&& 
\rho(r)=\rho_c +{\rho_2\over 2}r^2+ ... ,\;
p(r)= p_c+\frac{p_2}{2} r^2+... ,
\nonumber
\end{eqnarray}
where $\chi_c$,  $\rho_c$ and $p_c$ 
are the central values of the scalar field, the energy density and 
the pressure, respectively.
Note that using the scaling freedom of the $t$ coordinate, we set
 $N(0)=1$.
From Eqs.~(\ref{fe_B})--(\ref{eq_chi}), we obtain
\begin{eqnarray}
3 B_2&=&-8\pi \hat{G}\rho_c -\chi_c V_c -3C_2,
\\
B_2+2N_2& =& 8\pi \hat{G}p_c-\chi_c V_c-2C_2,
\\
3C_2& =& \frac{8\pi \hat{G}}{3}(-\rho_c+3p_c)+\frac{2\chi_c^2}{3}V_{\chi_c},
\end{eqnarray}
where $\hat{G}:=G/\chi_c$ is the effective gravitational constant,
$V_c:=V(\chi_c)$, and $V_{\chi_c}=dV/d\chi|_{\chi=\chi_c}$.
These three equations are rearranged to give
\begin{eqnarray}
B_2&=&-\frac{8\pi \hat{G}}{9}\left(2\rho_c+3 p_c\right)
-\frac{\chi_c}{3}V_c-\frac{2\chi_c^2}{9}V_{\chi_c},\label{ex-B2}
\\
N_2&=&\frac{8\pi \hat{G}}{9}(2\rho_c+3p_c)-\frac{\chi_c}{3}V_c
-\frac{\chi^2_c}{9}V_{\chi_c},
\\
C_2&=&\frac{8\pi\hat{ G}}{9}(-\rho_c+3p_c)+\frac{2\chi_c^2}{9}V_{\chi_c}.
\end{eqnarray}
Then, $p_2$ is derived from the conservation equation:
\begin{eqnarray}
p_2+N_2\left(\rho_c+p_c\right)=0.\label{ex-p2}
\end{eqnarray}
The Ricci scalar is given by
$R= R_c+{\cal O}(r^2)$ with $R_c=-6(B_2+N_2)$ near $r=0$.

If the energy density is constant inside the star, $\rho=\rho_0$,
Eq.~(\ref{dp=}) immediately gives
\begin{eqnarray}
N(r) = \left[\frac{\rho_0+p_c}{\rho_0+p(r)}\right]^2.
\end{eqnarray}
In the rest of the paper, we focus on constant density stars for simplicity.

\subsection{A classical mechanics picture}

\begin{figure}[tb]
  \begin{center}
    \includegraphics[keepaspectratio=true,height=55mm]{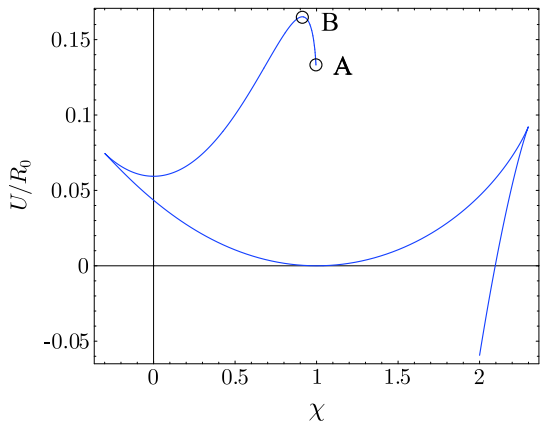}
  \end{center}
  \caption{The (inverted) potential $U(\chi)$ for Starobinsky's $f(R)$ model with $\lambda=2$ and $n=1$.
  The point A corresponds to a curvature singularity $(R=+\infty)$, and the point B is
  the de Sitter extremum.
  (See also Fig.~1 of Ref.~\cite{Frolov}.)}%
  \label{fig:u1.eps}
\end{figure}

\begin{figure}[tb]
  \begin{center}
    \includegraphics[keepaspectratio=true,height=50mm]{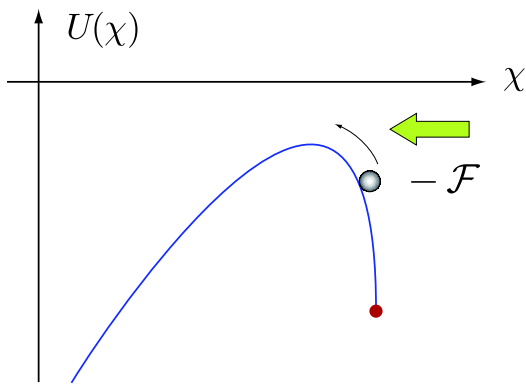}
  \end{center}
  \caption{Motion of a particle near the de Sitter extremum of $U(\chi)$.
  The particle feels the force ${\cal F}$ ($<0$) which arises from the trace of
  the energy-momentum tensor of the matter, ${\cal F}\propto T$.}%
  \label{fig:u2.eps}
\end{figure}

Given $\rho_0$, $p_c$, and $\chi_c$ (or, equivalently, $R_c$),
Eqs.~(\ref{dp=})--(\ref{eq_chi}) can be integrated outwards from the center
 to the
surface of the star, $r={\cal R}$, which is defined by $p({\cal R})=0$.
Then, one integrates the vacuum field equations from the star surface
to sufficiently large $r$, finding the exterior profile of the metric and
 the scalar field.
Unlike in general relativity, we have the extra scalar degree of freedom
corresponding to the choice of $\chi_c$.
However, not all values of $\chi_c$ can lead to a physically reasonable
solution inside and outside a star. The desired solution is such that
$\chi(r)\to\chi_1$ as $r\to\infty$, i.e., the asymptotically 
de Sitter solution with
the cosmological constant $\Lambda_{{\rm eff}}$.

We can formulate the problem of finding the physical configuration of 
stars as follows.
Eq.~(\ref{eq_chi}) can be written as
\begin{eqnarray}
\frac{d^2}{dr^2} \chi+\frac{2}{r}\frac{d}{dr}\chi= -\frac{dU}{d\chi}+{\cal F},
\label{dynamics}
\end{eqnarray}
where
\begin{eqnarray}
\frac{dU}{d\chi}:=\frac{1}{3}\left[ f_R Q-2f \right]=
-\frac{2\chi^3}{3}\frac{dV}{d\chi}
\label{dU}
\end{eqnarray}
and
\begin{eqnarray}
{\cal F}:=-\frac{8\pi G}{3}(\rho-3p).
\end{eqnarray}
Here we neglect the effect of the metric 
for the moment to comprehend the essential point.
[Later we will solve the full set of equations~(\ref{dp=})--(\ref{eq_chi}) 
numerically.]

Regarding $r$ as a time coordinate,
we find that  Eq.~(\ref{dynamics}) is a ``dynamical" equation
describing the motion of a particle in the potential $U$ under 
the time-dependent force ${\cal F}$.
The second term in the left hand side of Eq.~(\ref{dynamics}) represents 
frictional force,
which may also affect the dynamics in some cases.
The potential for the particle $U(\chi)$, defined by Eq.~(\ref{dU}),
is different from the inverted potential $-V(\chi)$.
However, the structure of $U$ around $\chi=\chi_1$ is quite similar to $-V$, 
as is shown in Figs.~\ref{fig:u1.eps} and~\ref{fig:u2.eps}:
the point A ($\chi=1$) corresponds to a curvature singularity, $R\to\infty$,
and the point B ($\chi=\chi_1$) is the de Sitter extremum.

In Ref.~\cite{Frolov}, Frolov similarly introduced the potential $-U(\chi)$
and the force term which is essentially given by the trace of 
the energy-momentum tensor.
For the purpose of solving for the radial profile, it is more convenient
to consider the inverted potential $-(-U) = U$,
as in the cases of  bubble nucleation~\cite{bubble} and of the chameleon 
model~\cite{Chameleon}

Suppose that the initial position of the particle, $\chi_c$, lies between 
points A and B. 
The particle starts at rest since $\chi'|_{r=0}=0$.
The force term ${\cal F}$
depends on the matter configuration inside the star and
plays a crucial role in this problem.
When $\rho>3p$, we have ${\cal F}<0$.
Since the pressure becomes smaller for larger $r$,
the force $|{\cal F}|$ is stronger near the surface than in the central region
 of the star. 
The force vanishes for $r>{\cal R}$ (if one assumes the vacuum exterior).

For fixed $\rho_0$ and $p_c$,
the behavior of the particle depends on its initial position.
If $\chi_c$ is sufficiently close to $1$,
the slope of the potential is bigger than the force term ${\cal F}$ initially,
so that $\chi$ will rapidly roll down to the curvature singularity, $\chi=1$.
Let $\chi_{s}$ be the minimum value of $\chi_c$ for which this occurs.
Specifically, $\chi_s$ is found by solving the equation
 $[{\cal F}-dU/d\chi]_{\chi_s}=0$.
If $\chi_c<\chi_s$, the particle initially
climbs up the potential hill under the force ${\cal F}<0$.
The force vanishes at $r={\cal R}$, but
the particle keeps climbing up the potential for the moment.
For $r>{\cal R}$, the scenario one can easily deduce is as follows:
The particle
cannot reach the top of the potential hill and turns around at some $r$,
ending up with the curvature singularity $\chi=1$,
or goes through the top and rolls down to the left, depending on
the initial position $\chi_c$~\footnote{The
particle which climbs up the potential initially never turns around
before $r={\cal R}$. This is because $|{\cal F}|\propto\rho_0-3p$
never decreases inside the star, and
the potential is less steep for smaller $\chi$.
This will be true
for more realistic profiles of the density and pressure,
except in the vicinity of the star surface.
Note, however, that it is possible for $\chi$ to pass through $\chi_1$ before 
$r={\cal R}$.
}.
In this case, one finds a critical value $\chi_{{\rm crit}}$ between the 
``turn-around''
and ``rolling-down'' solutions. By fine-tuning the initial position so that 
$\chi_c=\chi_{{\rm crit}}$,
we can realize the asymptotically de Sitter solution for which $\chi\to\chi_1$
 as $r\to\infty$.
Thus, the problem reduces to a boundary value problem.

However, there is another possibility that the particle inevitably overshoots 
the potential
even for the possible maximum value of $\chi_c$ (i.e., $\chi_s$).
In this case, one cannot obtain the desired solution:
the particle rolls down into the curvature singularity right after it starts 
to move,
or overshoots the potential.
Since the gravitational potential produced by a star is proportional to
$(\rho_0{\cal R}^3)/{\cal R} = \rho_0{\cal R}^2$,
a stronger gravitational field implies stronger force and/or
a longer period during which the force term survives effectively.
As a result, the fine-tuned initial location $\chi_{{\rm crit}}$
goes toward the right
as the star accommodates a larger gravitational potential,
and $\chi_{{\rm crit}}$ will eventually reach the point $\chi_s$.
Therefore, it is expected that
there is a maximum value of the gravitational potential for
a star to exist. 



\section{Analytic argument}\label{sec:An}

Before solving the full
 set of equations~(\ref{dp=})--(\ref{eq_chi}) 
 numerically,
in this section we shall provide some analytic arguments.
The analysis with approximate solutions
will help to understand
our numerical results presented in Sec.~\ref{sec:Num}.

First let us consider the interior of a star: $r<{\cal R}$.
To give a tractable argument, we assume that
\begin{eqnarray}
|B-1|,\;|N-1|\ll 1,\quad |B'/B|,\;|N'/N|\ll r^{-1}.
\end{eqnarray}
Then, the solution to Eq.~(\ref{eq_chi}) which is regular at the center 
is given by 
\begin{eqnarray}
\chi'\simeq- \frac{2G\mu}{3{\cal R}^3}r,\quad
\chi\simeq\chi_c-  \frac{G\mu}{3{\cal R}^3}r^2,
\label{apr_sol_chi}
\end{eqnarray}
where we have defined
\begin{eqnarray}
\mu :=\frac{4\pi}{3} \left(\rho_0 -\frac{\chi^3_c}{4\pi G} 
V_{\chi_c} \right){\cal R}^3.
\end{eqnarray}
Here we have ignored the pressure $p$ relative to $\rho_0$,
and made a rough approximation $\chi^3V_\chi\approx \chi_c^3V_{\chi_c}$.
Using Eqs.~(\ref{fe_B}) and~(\ref{apr_sol_chi}),
 we obtain
\begin{eqnarray}
B\simeq1-\frac{2\hat{G} (M-\mu/3)}{ {\cal R}^3} r^2 ,
\label{bin}
\end{eqnarray}
where $M:=4\pi \rho_0{\cal R}^3/3$.
Here we have neglected the ``cosmological 
constant'' $\chi^2 V$.
(We remind the reader that $\hat G:=G/\chi_c$.)

To derive the exterior solution, we approximate
\begin{eqnarray}
\frac{2\chi^3}{3}\frac{dV}{d\chi}\simeq \frac{\chi-\chi_1}{\lambda_\chi^2},
\end{eqnarray}
where $\lambda_\chi^{-2}:=({2\chi_1^3}/{3})\,{d^2V}/{d\chi^2}|_{\chi_1}$,
and analyze the behavior of $\chi$ around $\chi_1$.
The exterior solution to Eq.~(\ref{eq_chi}) is found to be
\begin{eqnarray}
\chi\simeq \chi_1+{\cal C}\,\frac{e^{-(r-{\cal R})/\lambda_\chi}}{r}.
\label{chiout}
\end{eqnarray}
Matching this to the interior solution~(\ref{apr_sol_chi}) at $r={\cal R}$, 
we obtain
\begin{eqnarray}
{\cal C}&\simeq&\frac{2G\mu}{3},
\\
\chi_c&\simeq&\chi_1+\frac{G\mu}{{\cal R}},
\label{thin-s}
\end{eqnarray}
where we have used 
$\lambda_{\chi}\approx{\cal O}(\Lambda_{{\rm eff}}^{-1/2})\gg{\cal R}$.
Given $\rho_0$, Eq.~(\ref{thin-s}) determines $\chi_c$.
Solving Eq.~(\ref{fe_B}) and
 matching the solution to the interior one~(\ref{bin}) at the surface of 
the star, we find
\begin{eqnarray}
B\simeq 1- \frac{2\hat{G}(M- \mu/3 )}{r}.\label{bout}
\end{eqnarray}
Then, Eq.~(\ref{fe_N}) implies that
\begin{eqnarray}
N\simeq N_\infty\left[1- \frac{2\hat{G}(M+ \mu/3) }{r}\right]
\,,
\label{nout}
\end{eqnarray}
where the asymptotic value $N_\infty $ is not unity
because of our boundary condition at the center, but it
can be set unity by rescaling of the time coordinate.
From Eqs.~(\ref{bout}) and~(\ref{nout}),
the PPN parameter $\gamma$ is found to be
\begin{eqnarray}
\gamma=\frac{3M-\mu}{3M+\mu}.
\end{eqnarray}

Now let us define the gravitational potential evaluated at the surface of
 the star:
\begin{eqnarray}
\Phi:=\frac{\hat{G}(M+\mu/3)}{{\cal R}}.
\end{eqnarray}
In terms of this, Eq.~(\ref{thin-s}) can be written as
\begin{eqnarray}
\frac{\Delta}{\Phi} = \frac{3\mu}{3M+\mu} \quad\text{with}\quad 
\Delta:=\frac{\chi_c-\chi_1}{\chi_c}.
\end{eqnarray}
The ``thin-shell'' condition is given by $\Delta\ll \Phi$~\cite{Chameleon}. 
This is equivalent to $\mu\ll M$, which leads to ${\cal C}\ll GM$ and
indeed suppresses the deviation of $\chi(r)$ from $\chi_1$ outside the star.
This situation is realized if $4\pi G\rho_0- \chi_c^3V_{\chi_c}\approx 0$.
On the other hand, if $4\pi G\rho_0\gg \chi_c^3V_{\chi_c}$, and hence 
$\mu\simeq M$,
the thin-shell condition does not hold.
It is easy to see that 
$\gamma\simeq 1/2$.
In this ``thick-shell'' case, $\chi_c$ must be not too far from
$\chi_1$ so as not to fall into the curvature singularity, $\chi=1$.
Therefore, we require that $\chi_c<1$.
This condition together with Eq.~(\ref{thin-s}) gives the bound
\begin{eqnarray}
\Phi < \Phi_{{\rm max}} =\frac{4}{3}(1-\chi_1),
\end{eqnarray}
i.e., stars with $\Phi> \Phi_{{\rm max}}$ cannot exist.
As shown later by numerical solutions, the thin-shell condition is violated
as long as the exterior is vacuum.

Specifically, for $n=1$ one needs $x_1>\sqrt{3}$ in order to have a 
de Sitter minimum.
This leads to $1-\chi_1<1/3$ ($\Phi_{{\rm max}}<4/9$).
For $x_1=3.6$ ($\lambda=2.088$), we have $1-\chi_1=0.07716$ ($\Phi_{{\rm max}}=0.1029$).
For $n=2$, one needs $x_1>\sqrt{\sqrt{13}-2}$, giving $1-\chi_1\lesssim 0.2705$
($\Phi_{{\rm max}}\lesssim 0.3606$).

\section{Numerical results}
\label{sec:Num}

\subsection{Stars in $f(R)$ gravity}

\begin{figure}[tb]
  \begin{center}
    \includegraphics[keepaspectratio=true,height=52mm]{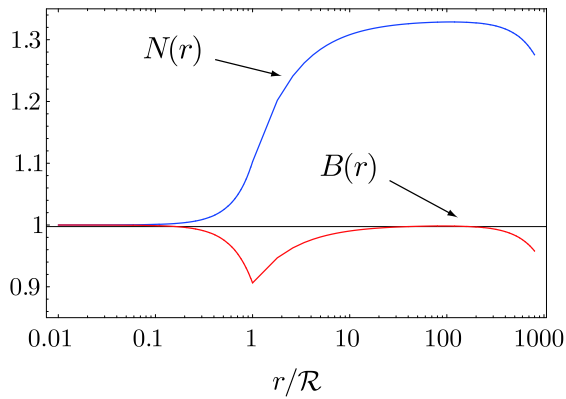}
  \end{center}
  \caption{Metric for a nonrelativistic star. Parameters are given by
  $n=1$, $x_1=3.6$, $4\pi G\rho_0=10^6\Lambda_{{\rm eff}}$, and $p_c=5\times10^{-2}\rho_0$.
  The central value of the Ricci scalar is tuned to be $R_c = 3.462\times 10^{-6}\times8\pi G \rho_0$.
  The radial coordinate is normalized by the radius of the star ${\cal R}$.}%
  \label{fig:metric1.eps}
\end{figure}

\begin{figure}[tb]
  \begin{center}
    \includegraphics[keepaspectratio=true,height=51mm]{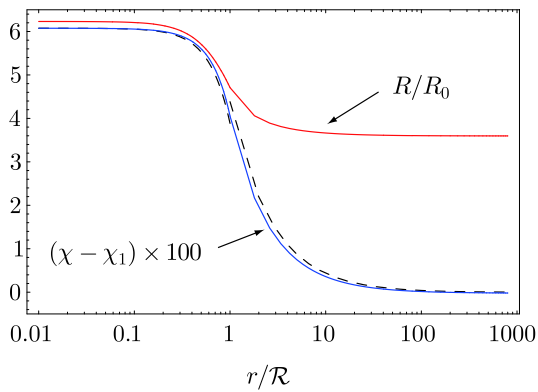}
  \end{center}
  \caption{Numerical solutions of the Ricci scalar and $\chi$ for 
a nonrelativistic star. Parameters are the same as those in Fig.~\ref{fig:metric1.eps}.
  Dashed line is a plot of the analytic approximation~(\ref{apr_sol_chi}) 
and~(\ref{chiout}). Since $R\to x_1R_0$ and $\chi\to\chi_1$ as $r\to\infty$,
this is the desired solution with asymptotically de Sitter geometry.}%
  \label{fig:curvature1.eps}
\end{figure}

\begin{figure}[tb]
  \begin{center}
    \includegraphics[keepaspectratio=true,height=51mm]{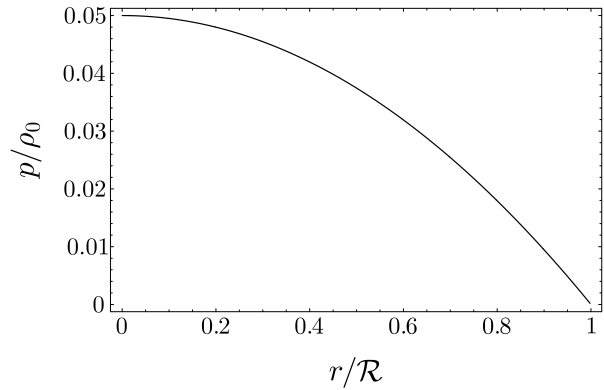}
  \end{center}
  \caption{The pressure profile $p(r)$ of a nonrelativistic star.
  Parameters are the same as those in Fig.~\ref{fig:metric1.eps}.}%
  \label{fig:pressure1.eps}
\end{figure}

We numerically integrate Eqs.~(\ref{dp=})--(\ref{eq_chi}) outwards
from  the center, imposing the appropriate boundary conditions
given by Eq.~(\ref{series-r=0}) with Eqs.~(\ref{ex-B2})--(\ref{ex-p2}).
We reconstruct $R=R(r)$ from the metric and compare
it with the solution for the scalar degree of freedom, $R=Q(\chi(r))$,
to make sure that numerical errors are sufficiently small.
The angular component of the field equations is also used for the same purpose.


The sets of the model parameters we use are:
(i) $n=1$ and $x_1=3.6\;(\lambda =2.088)$, for which $\chi_1=0.9228$;
(ii) $n=2$ and $x_1=3.6 \;(\lambda =1.827)$, for which $\chi_1=0.9903$.
For numerical solutions
we shall take $4\pi G\rho_0=10^6\Lambda_{{\rm eff}}$.
This is not a realistic value, e.g., for neutron stars, but
we do not have to be concerned about this point because
the properties of stellar solutions which we
are interested in are basically characterized by
the gravitational potential rather than the energy density itself,
provided that $G\rho_0\gg\Lambda_{{\rm eff}}$.
This is expected from the analytic result, and we have confirmed that it
 is indeed true for
our numerical solutions.

\subsubsection{$n=1$ and $x_1=3.6$}

Taking $4\pi G\rho_0=10^6\Lambda_{{\rm eff}}$ and $p_c=10^{-4}\rho_0$,
we find an asymptotically de Sitter solution of a star for 
$R_c=2.001\times 10^{-6}\times 8\pi G\rho_0$.
The solution agrees well with the analytic approximation 
in Sec.~\ref{sec:An}.

A numerical calculation has been performed also in the case of
$4\pi G\rho_0=10^6\Lambda_{{\rm eff}}$ and $p_c=5\times10^{-2}\rho_0$.
We find a solution
by tuning $R_c = 3.462\times 10^{-6}\times8\pi G \rho_0$
($\chi_c=0.9836$), for which
\begin{eqnarray}
\frac{4\pi}{3}\hat{G}\rho_0{\cal R}^2=\frac{\hat G M}{{\cal R}}\simeq0.06687.
\label{grav_potential}
\end{eqnarray}
Our numerical result is shown in 
Figs.~\ref{fig:metric1.eps}--\ref{fig:pressure1.eps}.
One can see from Fig.~\ref{fig:curvature1.eps} that
$R\to x_1R_0$ and $\chi\to \chi_1$ as $r\to\infty$.
Since $\Delta=0.06176$ is almost the
 same as the gravitational potential
(\ref{grav_potential}),
the thin-shell does not form.
A numerical fitting leads to the approximate expression
for the exterior metric:
\begin{eqnarray}
N&\simeq& N_\infty\left(1-2c_1 \frac{{\cal R}}{r}
-\frac{c_2}{3}\Lambda_{{\rm eff}}r^2\right),
\\
B&\simeq&  1-2c_3 \frac{{\cal R}}{r}-\frac{c_4}{3}\Lambda_{{\rm eff}}r^2,
\end{eqnarray}
with
\begin{eqnarray}
N_{\infty}=1.332, \;
c_1=0.08716,\;
c_2=0.9973,\;
\nonumber\\
c_3=0.04747,\;
c_4=0.9993.
\end{eqnarray}
The PPN parameter turns out to be $\gamma\simeq c_3/c_1\simeq 0.5446$,
which is close to $1/2$, as expected.

\begin{figure}[tb]
  \begin{center}
    \includegraphics[keepaspectratio=true,height=52mm]{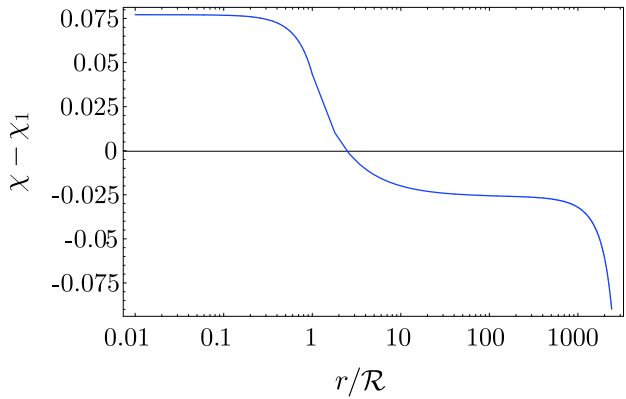}
  \end{center}
  \caption{A rolling-down solution for a would-be relativistic star.
  Parameters are given by $n=1$, $x_1=3.6$, $4\pi G\rho_0=10^6\Lambda_{{\rm eff}}$,
and $p_c=0.1\times\rho_0$. The central value of the Ricci scalar is $R_c=0.7000\times8\pi G\rho_0$.
The solution clearly overshoots the de Sitter extremum, $\chi=\chi_1$.}%
  \label{fig:chi_roll.eps}
\end{figure}

To explore stars with larger $\hat G M/{\cal R}$,
we have tried to find numerical solutions
for $4\pi G\rho_0=10^6\Lambda_{{\rm eff}}$
and $p_c=0.1\times\rho_0$.
A solution which is regular inside the star is obtained, e.g., for 
$R_c=0.7000\times8\pi G\rho_0$,
but this is the rolling-down solution (see Fig.~\ref{fig:chi_roll.eps}).
Hence it is unphysical
($N(r)\to 0$ and $B(r)\to\infty$ as $r\to\infty$),
though the gravitational potential is as ``large'' as
\begin{eqnarray}
\frac{\hat G M}{{\cal R}}\simeq0.1203.
\end{eqnarray}
Taking a slightly larger value of the central curvature,
$R_c=0.7001\times 8\pi G\rho_0$,
the Ricci scalar rapidly diverges inside the star.
This corresponds to the case in which
$\chi$ starts to move to the right toward the curvature singularity, $\chi=1$.
Note that since $\chi^3V\approx R/2$, $R_c=0.7\times 8\pi G\rho_0$
implies $4\pi G(\rho_0-3p_c)-\chi_c^3V_{\chi_c}\approx0$.

\subsubsection{$n=2$ and $x_1=3.6$}

For
$4\pi G\rho_0=10^6\Lambda_{{\rm eff}}$ and $p_c=5\times 10^{-4}\rho_0$,
we find the stellar configuration with de Sitter asymptotic behavior
by tuning $R_c=2.035\times 10^{-6}\times 8\pi G\rho_0$ ($\chi_c=0.9911$).
In this case, $\hat G M/{\cal R} =7.491\times 10^{-4}$.
The thin-shell condition does not hold ($\Delta=7.448\times 10^{-4}$),
and $\gamma=0.5005$.

Looking for stars with stronger gravitational fields, we take
$4\pi G\rho_0=10^6\Lambda_{{\rm eff}}$ and $p_c= 10^{-2}\rho_0$.
When $R_c=0.9700\times 8\pi G\rho_0$, a rolling-down solution is found
with $\hat G M/{\cal R}=0.01465$, while a slightly larger value $R_c=0.9701\times 8\pi G\rho_0$
leads to a curvature singularity $R\to\infty$ inside the star.

\vspace{8mm}

From the above numerical analysis, we conclude that
stars with strong gravitational fields cannot be present
in this class of $f(R)$ theories.

\subsection{Stars in surrounding media}

So far we have examined the case of the vacuum exterior.
Although we can indeed construct a stellar configuration provided that
gravity is weak, the exterior metric is not given by
the de Sitter-Schwarzschild solution in general relativity and
the PPN parameter is found to be $\gamma\simeq1/2$.
This simply reflects the fact that the thin-shell condition is violated in the present case.
However,
we can make the chameleon mechanism effective
by taking into account the effect of surrounding media (e.g., dark matter).
Let us make a brief comment on this point.

In a ``realistic'' situation, exterior matter is present around a star,
giving rise to the force term ${\cal F}_{{\rm ext}}\simeq-8\pi G\rho_{{\rm ext}}/3$ there.
Then, to obtain a viable stellar configuration, one has to take a shot
at the point $\chi_*$ satisfying the equation 
${\cal F}_{{\rm ext}}-dU/d\chi= 0$ rather than
the top of the potential hill, $\chi_1$.
Since $\chi_*>\chi_1$, the difference in $\chi$
between inside and outside the star
becomes smaller, and hence
it is easier to satisfy the thin-shell condition.
Indeed, the chameleon mechanism has been shown to work
in the $f(R)$ model which is very similar to the current one,
reproducing $\gamma\simeq 1$ in the solar vicinity~\cite{Hu}.


The above argument, however, only applies to stars with weak gravitational 
fields.
When gravity is strong, 
for any initial condition of $\chi_c$ which avoids 
the curvature singularity,
the scalar field $\chi$
inevitably overshoots the potential. It is clear from this fact that one 
cannot stop $\chi$
at any value of  $\chi_*>\chi_1$.
Therefore, 
a relativistic star cannot be present even with surrounding medium.

\section{Conclusions}\label{sec:Conc}

In this paper, we have studied
the strong gravity aspect
of $f(R)$ modified gravity models that
reproduce the conventional cosmological evolution and
evade solar system and laboratory tests~\cite{St, Hu, AB}.
It is known that $f(R)$ theories can be recasted simply in
the Brans-Dicke theory with $\omega=0$,
but the potential for the effective scalar degree of freedom
may play a complicated and nontrivial role.
Moreover, the presence of matter may affect dynamics of the scalar field,
possibly mimicking the chameleon model~\cite{Chameleon}.

We have explored
uniform density, spherically symmetric stars and their exterior geometry
in the $f(R)$ model of~\cite{St}.
The main result of the present paper is
summarized as follows: given model parameters, there is a maximum value
of the gravitational potential produced by a star, above which
no asymptotically de Sitter stellar configurations can be constructed.
We show this both analytically and numerically.
For example, the model with $n=1$ and $\lambda\approx 2$ gives 
$\Phi_{{\rm max}}\approx 0.1$.
This raises a warning sign for a class of $f(R)$ theories,
because neutron stars cannot be present in such gravity models.

The underlying mechanism that hinders strong gravitational fields around matter
is explained essentially as follows~\cite{Frolov}.
Consider a static matter distribution. The Newtonian potential obeys
the Poisson equation $\nabla^2\Phi \sim G \rho$,
while the equation of motion for the scalar field implies
$\nabla^2\chi \sim G\rho$.
From this, one can evaluate
 the excitation of the scalar degree of freedom around
the matter distribution
as $\delta\chi\sim{\cal O}(\Phi)$.
If the de Sitter minimum is located very close to the point 
$\chi=1$, which
corresponds to $R=\infty$ in the effective potential,
a slightly strong gravitational field will cause the problem of 
appearance of a curvature 
singularity.

Bearing the above evaluation in mind,
let us comment on the other specific models of $f(R)$ gravity.
The model of Hu and Sawicki~\cite{Hu}
and Starobinsky's one
share the
same structure of $f(R)$ in the high-curvature regime, i.e.,
$f(R)\approx R-2\Lambda_{{\rm eff}}+C/R^{\alpha}$ with $\alpha>0$.
Therefore, we expect that the same problem arises
in the Hu and Sawicki's model.
The model by Appleby and Battye is characterized by~\cite{AB}
\begin{eqnarray}
f(R)=\frac{R}{2}+\frac{1}{2a}\ln\left[
\cosh(aR)-\tanh(b)\sinh(aR)
\right],
\end{eqnarray}
where $a$ and $b$ are parameters.
Since $\chi = df/dR = [1+\tanh(aR-b)]/2$, a positive curvature 
singularity corresponds to $\chi=1$.
Taking, for example,
$b=1.5$, 
we find that $\chi_1\approx 0.93$ at the de Sitter minimum, 
which is very close to the dangerous curvature singularity
(the result is independent of $a$).
Since also in this model the effective potential is finite at $R=+\infty$,
we anticipate the same singularity problem.

Our choice of the parameters in the present paper gave $\Phi_{{\rm max}}\approx 0.1$,
for which neutron stars are unlikely to exist.
However, there still remains a possibility that
more realistic stellar environments and matter profiles
weaken the bound on the potential by a factor of 2 or 3,
and at the same time make the chameleon mechanism work~\footnote{
The neutron star-white
dwarf system PSR J1141--6545
can put strong
constrains on alternative theories of gravity around relativistic
 stars~\cite{Will, E-F}.
For this reason, we need to invoke the chameleon mechanism to describe
such stars.}.
It is technically much more difficult to construct stellar configurations
with a realistic equation of state,
realistic energy densities, and realistic stellar environments.
Such an elaborated modeling of relativistic stars might 
allow for $\Phi_{{\rm max}}$
as large as, say, 0.3, but
the parameter space of the theory will be very restricted.
To conclude, $f(R)$ theories that reproduce the correct behavior 
of weak gravity
in the solar vicinity do not admit neutron star solutions without 
special care.

\acknowledgments

T.K. would like to thank
 Kenta~Kiuchi for valuable discussions on neutron stars.
This work was partially supported by the JSPS under Contact No.~19-4199,
by the Grant-in-Aid for Scientific Research
Fund of the JSPS (No.~19540308) and by the
Japan-U.K. Research Cooperative Program.


\end{document}